\documentclass[twocolumn,showpacs,preprintnumbers,amsmath,amssymb]{revtex4}
\usepackage{graphicx}
\usepackage{dcolumn}
\usepackage{bm}

\begin{document}

\title{Atomic Scale Fractal Dimensionality in Proteins}

\author{Duccio Medini}
\affiliation{Immunobiological Research Institute of Siena, Chiron Vaccines,
Siena, Italy}
\author{A. Widom}
\affiliation{Physics Department, Northeastern University, Boston MA USA}

\date{\today}

\begin{abstract}
The soft condensed matter of biological organisms exhibits 
atomic motions whose properties depend strongly on temperature 
and hydration conditions.  Due to the 
superposition of rapidly fluctuating alternative motions at both very 
low temperatures (quantum effects) and very high temperatures 
(classical Brownian motion regime), the dimension of an atomic ``path'' 
is in reality different from unity. In the intermediate temperature 
regime and under environmental conditions which sustain active 
biological functions, the fractal dimension of the sets upon which 
atoms reside is an open question. Measured values 
of the fractal dimension of the sets on which the Hydrogen 
atoms reside within the Azurin protein
macromolecule are reported. The distribution of proton positions 
was measured employing thermal neutron elastic scattering 
from Azurin protein targets. As the temperature was raised from low 
to intermediate values, a previously known and 
biologically relevant dynamical transition 
was verified for the Azurin protein only under hydrated conditions. 
The measured fractal dimension of the geometrical sets on which protons 
reside in the biologically relevant temperature regime is given by 
$D=0.65 \pm 0.1$. The relationship between fractal dimensionality 
and biological function is qualitatively discussed.
\end{abstract}

\maketitle

\section{Introduction}

There has been considerable interest in the fractal nature 
of atomic distributions within amorphous condensed 
matter\cite{Martin87,Orbach86} and in particular soft condensed 
matter such as proteins\cite{Wagner85,Colvin85,Elber86}. 
Fractal sets have been studied experimentally using 
X-ray and small angle neutron scattering techniques which 
probe the correlations between different atoms within the target. 
Geometrical based numerical calculations \cite{Wang90,Xiao94} 
have been used to determine the fractal dimension 
of secondary and tertiary structures of 90 proteins.
For Azurin, the secondary structure of 
\begin{math}\beta \end{math}-sheets and reverse-turns 
had the computed fractal dimension 
\begin{math}D_1 = 1.35 \pm 0.04\end{math}.  
The tertiary structure of global folding 
had the computed fractal dimension 
\begin{math}D_2 = 1.68 \pm 0.08\end{math}.   
In the above cases, coherence lengths 
\begin{math} \xi \end{math} 
of the order of the secondary 
(\begin{math} \xi \sim 10\end{math}\AA)  
and the tertiary structure 
(\begin{math} \xi \sim 40 \end{math}\AA) 
were of interest. 

Rarely can the motion of an atom within biological macromolecules 
be described as a simple one-dimensional path.
The soft condensed matter of biological organisms exhibits atomic
motions whose properties depend strongly on 
temperature\cite{Parak71,Smith91} and hydration 
conditions\cite{Ferrand93,Reat00}. Living organisms 
themselves are active only within a quite limited range of
temperatures and under particular environmental conditions. For very low
and very high temperatures, respectively, the entropy of soft condensed
biological matter is either too small (almost perfect order) or too
large (severe disorder) to sustain normal biological functions. 
One characterization of the intermediate range of temperatures required for
life employs the fractal dimension of the paths traversed by atoms within
biological matter\cite{Orbach86,Wagner85,Entin-Wohlman89,Elber90,Teixeira86}. 
It is this intermediate range of temperature that is associated with 
a well known\cite{Parak71,Frauenfelder79,Smith91,Paciaroni00} and 
biologically relevant\cite{Parak80,Lehnert93} dynamical transition from 
almost harmonic oscillations to strongly anharmonic modes of motion.   

Ordinary paths are usually thought to be one-dimensional sets. 
However, at very low temperatures, quantum mechanics smears out 
the notion of a path; i.e. there exists quantum mechanical superposition 
of amplitudes of alternative paths. At sufficiently high temperatures 
there is superposition of Einstein-Brownian motion 
probabilities\cite{Mandelbrot82,Orbach86}. A normal path
described by a velocity \begin{math} v  \end{math} with physical units 
of \begin{math} [cm/sec] \end{math} is one-dimensional. A 
Brownian motion ``path'' described by a diffusion
coefficient \begin{math} D \end{math} with physical units 
of \begin{math} [cm^2/sec] \end{math} 
or area per unit time is in reality two-dimensional\cite{Mandelbrot82}. 
In the intermediate temperature range of living organisms, biomedical 
proton spin nuclear magnetic imaging methods exhibit an apparent 
diffusion coefficient (ADC) varying in time
indicating fractal dimensional diffusion paths for the measured 
protons\cite{Widom95}. The fractal path dynamics of atoms will be manifest 
in the fractal dimensions of the spatial sets on which the atoms reside.
In the intermediate temperature regime and only under environmental 
conditions which sustain active biological functions, the fractal 
dimension of these spatial sets is an open 
question\cite{Orbach86,Wagner85}. 

Our purpose is to present data from elastic neutron scattering off 
protons (hydrogen atomic nuclei) within the Azurin protein macromolecule. 
To our knowledge no high momentum-exchange incoherent neutron 
investigations have been employed to discuss the fractal nature
of the proton distributions in proteins. The length scales here 
are truly microscopic 
(of order \begin{math} \xi \sim 1\end{math}\AA).
Most importantly, none of the previous studies involves a discussion 
of the temperature dependence of the fractal exponents of the proton 
distributions in the dynamical-transition region.
In our experiments, when temperature was raised from low to 
intermediate values, the previously known dynamical 
transition\cite{Smith91,Paciaroni00} was verified for 
the Azurin protein only under hydrated conditions. The resulting data 
substantiates that Hydrogen atomic protons reside on fractal geometrical 
sets of dimension \begin{math} D=(0.65 \pm 0.1) \end{math} but 
only in the intermediate biologically relevant temperature regimes and 
only in the presence of an amount of water required for biological 
activity. 

In Sec.2, it is shown how the sets on which atoms 
reside may be deduced from neutron scattering data. The physical 
pictures which appear from such  neutron data are compared with 
those which have been typically obtained from X-ray scattering. 
In Sec.3, the mathematical expression for 
\begin{math} D \end{math}-dimensional fractal form factors is derived. 
The more conventional Debye Waller expression is shown to be a special 
limiting case of a fractal form factor. 
In Sec.4, the experimental methods for measuring the neutron 
elastic scattering cross section are discussed in some detail. 
The data is presented in Sec.5. In the concluding Sec.6, 
the relationships between fractal dimensionality and biological 
functions are qualitatively discussed.

\section{Scattering measurements of geometrical sets}
In order to detect the geometrical sets in space within which atoms
reside, it is useful to generate pictures of how the Azurin
macromolecule appears to a given experimental probe. The measured
fractal dimensionality of sets depends on the nature of the probe, 
the target and the length scale being 
investigated\cite{Wagner85,Martin87,Chen86}.
Two commonly used experimental diffraction probes for 
taking pictures of proteins are X-ray photons and thermal 
neutrons.

\begin{figure}[bp]
\scalebox {0.5}{\includegraphics{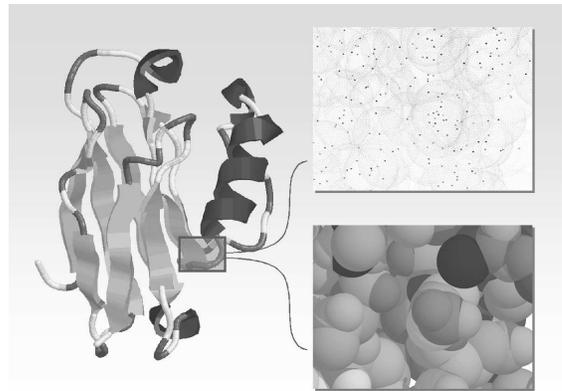}}
\caption{Shown is the globular Azurin protein involved in
electron-transfer processes. The protein structure (pdb-entry 1azu)
is here pictured for the second of the four monomers in the 
crystalographic complex. In the magnified regions of the figure 
are shown (i) the expected positions of protons to be 
probed by thermal neutron diffraction and (ii) the electron clouds 
of the atoms as detected by X-ray difraction. The localized proton 
distributions are superimposed on the electron clouds which are 
less localized in space.}
\label{Fig1}
\end{figure}

The main contribution to the elastic scattering of thermal neutrons 
off protein targets comes from the incoherent scattering from 
Hydrogen nuclei\cite{Lovesey84}. The elastic differential cross section for 
the neutron to scatter from 
\begin{math}N\end{math} Hydrogen atom protons within the protein 
macromolecule is then given by\cite{Lovesey84} 
\begin{equation}
\left( \frac{d\sigma}{d\Omega}\right)_{el} \approx 
\left( \frac{N \sigma_{inc}}{4 \pi}\right) 
\left| F({\bf Q}) \right|^2 ,
\label{cross_section}
\end{equation}
where \begin{math} \sigma_{inc}  \end{math} is the single proton 
incoherent neutron cross section and \begin{math} F({\bf Q}) \end{math} 
is the proton form factor corresponding to a neutron momentum transfer 
of \begin{math} \hbar{\bf Q}={\bf p}_i-{\bf p}_f \end{math}. In detail, 
if \begin{math} \rho({\bf r})d^3 {\bf r}\end{math} denotes the probability 
of finding a proton (within an Hydrogen atom) at position 
\begin{math}{\bf r} \in d^3 {\bf r}\end{math}, then the form factor 
\begin{math}F({\bf Q})\end{math} is defined as the Fourier transform  
\begin{equation}
F({\bf Q}) 
= \int \rho ({\bf r}) e^{-i {\bf Q} \cdot {\bf r}} d^3 {\bf r}.
\label{F_of_Q}
\end{equation}
Previous studies of fractal sets in 
proteins\cite{Lewis85,Teixeira86,Entin-Wohlman89,Elber90} 
relied mainly on protein structure so obtained by X-ray scattering 
data. The cross section for such 
scattering with a photon wave vector transfer of 
\begin{math} {\bf Q}={\bf k}_i-{\bf k}_f  \end{math} is given by 
\begin{equation}
\left( \frac{d\sigma}{d\Omega}\right)_{X-ray}=
\frac{1}{2}\left({e^2\over mc^2}\right)^2(1+\cos^2\Theta )
\overline{\left| f({\bf Q}) \right|^2}, 
\label{X_ray}
\end{equation}
and the X-ray form factor is given by the Fourier transform 
of the density of electrons \begin{math} n({\bf r}) \end{math}; i.e.   
\begin{equation}
f({\bf Q})=
\int n({\bf r}) e^{-i {\bf Q} \cdot {\bf r}} d^3 {\bf r}.
\label{e_density}
\end{equation}
The neutron scattering form factor in Eq.(\ref{F_of_Q}) probes  
the probability distribution in space 
\begin{math} \rho ({\bf r}) \end{math} of the protons 
while the X-ray scattering form factor probes the total mean 
electronic density \begin{math} \bar{n} \end{math} of all the 
atoms and the correlations 
\begin{math} C \end{math} between them; i.e.  
\begin{equation} 
\overline{n({\bf r}_1)n({\bf r}_2)}=\delta({\bf r}_1-{\bf r}_2)
\bar{n}({\bf r}_1)+\bar{n}({\bf r}_1)\bar{n}({\bf r}_2)
C({\bf r}_1,{\bf r}_2).  
\end{equation}

The constituents of the Azurin protein as seen by X-ray 
photons are shown in Fig.\ref{Fig1} using
experimental results stored in the Protein Data Bank at Brookhaven National
Laboratory\cite{Nar91}. The electronic densities of each atom are probed by X-ray 
diffraction. The expected picture of non-exchangeable Hydrogen atom proton
positions (deduced employing chemical structures from the same X-ray data)
are also shown in Fig.\ref{Fig1}. The non-exchangeable Hydrogen atom nuclei dominate
the picture of the Azurin macromolecule as probed by neutron diffraction on
atomic length scales. The electron density within an atom is represented 
in the picture by the empirical atomic radii\cite{Slater64}. 
The proton distributions are more localized than are the 
electron clouds about the nuclei. 

\section{Fractal form factors}
The usual (isotropic) Debye-Waller form factor\cite{Lovesey84}  
\begin{equation}
F_{Debye-Waller}({\bf Q})=e^{-Q^2\overline{|{\bf u}|^2}/6}
\label{3frac0a}
\end{equation}
corresponds to a Gaussian probability density 
\begin{equation}
\rho_{Debye-Waller}({\bf r})=
\left(\frac{3}{2\pi \overline{|{\bf u}|^2}}\right)^{3/2}
e^{-3|{\bf r}|^2/2\overline{|{\bf u}|^2}}.
\label{3frac0b}
\end{equation}
Such conventional form factors for Hydrogen atom nuclei 
assume harmonic oscillations in the position and are 
thereby valid only in the low temperature regime. For proteins 
at intermediate or high temperatures, one must take superpositions of 
Gaussian superpositions which may be employed to describe fractal form 
factors. How this comes about will now be discussed.

\begin{figure}[bp]
\scalebox {0.8}{\includegraphics{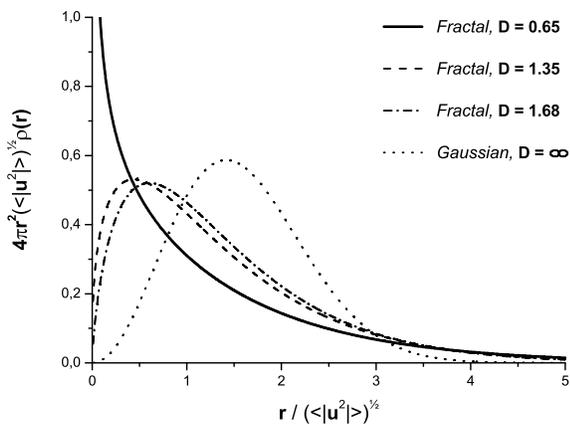}}
\caption{Shown is a plot of $(dP(r,D)/dr)=4\pi r^2 \rho_D(r)$ 
for the fractal dimensions describing (i) sets on which protons 
reside, (ii) secondary $\beta $-sheet structures, (iii) the tertiary 
folding structure and (iv) the conventional (Gaussian) Debye-Waller 
factor. The value of $\left<|{\bf u}|^2\right>$ is fixed.
Low values of $D$ imply a strong localization of the position 
distribution near the equilibrium site. The long ranged tail 
$\rho_D(r)$ for large $r$ inceases as $D$ grows smaller.}
\label{Fig2}
\end{figure}

If the position of a Hydrogen atom nucleus (as a random variable) 
is distributed isotropically in space, then the probability for such 
a proton to be found within a sphere of radius $R$ is given by 
\begin{equation}
P(R)=4\pi \int_0^R \rho (r)r^2 dr,\ \ {\rm where}\ \ 
\lim_{R\to \infty}P(R)=1.
\label{3frac1}
\end{equation}
The form factor in Eq.(\ref{F_of_Q}) may then be written 
\begin{equation}
F(Q)=4\pi \int_0^\infty \left(\frac{sin(Qr)}{Qr}\right)
\rho (r)r^2 dr.
\label{3frac2}
\end{equation}
A probability distribution \begin{math} P(R) \end{math} 
will be said to exhibit a fractal
dimension \begin{math} D \end{math} if and only if
\begin{equation}
P(R)\sim R^D\ \ {\rm as}\ \ R\to 0,
\label{3frac3}
\end{equation}
or equivalently if and only if
\begin{equation}
F(Q)\sim Q^{-D}\ \ {\rm as}\ \ Q\to \infty .
\label{3frac4}
\end{equation}
In practise, this implies the existence of a 
``coherence length'' \begin{math} \xi  \end{math} such that
Eq.(\ref{3frac3}) holds true in the regime 
\begin{math}R<<\xi \end{math} 
while Eq.(\ref{3frac4}) holds true in the regime 
\begin{math} Q>>1/\xi \end{math}.
A simple form factor\cite{Fisher67} incorporating both the notions 
of a fractal dimensionality \begin{math} D \end{math} and a 
coherence length  \begin{math} \xi \end{math} may be written as
\begin{equation}
F_D(Q)=\left(\frac{1}{1+(Q\xi)^2}\right)^{(D/2)}.
\label{3frac5}
\end{equation}
For all (isotropic) form factor models, the mean square 
displacement may be computed from\cite{Lovesey84}
\begin{equation}
\left<|{\bf u}|^2\right>=-6\lim_{Q^2\to 0}
\frac{d}{d(Q^2)}\ln F({\bf Q}).
\label{msd_definition}
\end{equation}
For form factor of Eq.(\ref{3frac5}), 
the coherence length \begin{math} \xi  \end{math} is related 
to the mean square displacement \begin{math} <|{\bf u}|^2> \end{math}
via 
\begin{equation}
\left<|{\bf u}|^2\right>=3D\xi^2 .
\label{3frac5a}
\end{equation}

The fractal form factor of Eq.(\ref{3frac5}) may alternatively be 
viewed as arising from an inhomogeneously  distributed ensemble 
of protons each with Gaussian distributions of varying widths. 
The mathematical proof of this interpretation will now be exhibited. 
The Gamma function for \begin{math}{\Re }e (z)>0 \end{math} is 
defined as
\begin{equation}
\Gamma (z)=\int_0^\infty e^{-y}y^z\left(\frac{dy}{y}\right).
\label{3frac6}
\end{equation}
Change variables in the above integral using 
\begin{math} y=as  \end{math} 
to obtain the identity
\begin{equation}
\left(\frac{1}{a}\right)^z=\frac{1}{\Gamma (z)}
\int_0^\infty e^{-as}s^z\left(\frac{ds}{s}\right).
\label{3frac7}
\end{equation}
Eqs.(\ref{3frac5}) and (\ref{3frac7}) for 
\begin{math} z=(D/2) \end{math} and 
\begin{math} a=1+(Q\xi)^2 \end{math} 
yield the form factor
\begin{equation}
F_D(Q)=\frac{1}{\Gamma (D/2)}
\int_0^\infty e^{-s}e^{-sQ^2\xi^2 }s^{(D/2)}\left(\frac{ds}{s}\right).
\label{3frac8}
\end{equation}                                                               

Let us now consider the density distribution due to a simple 
Gaussian form factor
\begin{eqnarray}
\rho_G({\bf r},s)&=&\int e^{-sQ^2\xi^2 }e^{i{\bf Q\cdot r}}
\left(\frac{d^3{\bf Q}}{(2\pi )^3}\right) \nonumber \\  
&=& \left(\frac{1}{4\pi s\xi^2}\right)^{(3/2)}e^{-|{\bf r}|^2/4s\xi^2}.
\label{3frac9}
\end{eqnarray}             
The fractal probability density
\begin{equation}
\rho_D({\bf r})=\int F_D(Q)e^{i{\bf Q\cdot r}}
\left(\frac{d^3{\bf Q}}{(2\pi )^3}\right)
\label{3frac10}
\end{equation}
may be written as a superposition of Gaussian densities  
\begin{equation}
\rho_D({\bf r})=\frac{1}{\Gamma (D/2)}
\int_0^\infty e^{-s}s^{(D/2)}\rho_G({\bf r},s)\left(\frac{ds}{s}\right).
\label{3frac11}
\end{equation}       
In detail, Eqs.(\ref{3frac8}), (\ref{3frac9}) 
and (\ref{3frac10}) imply that
\begin{eqnarray}
\rho_D({\bf r})=\frac{1}{\Gamma (D/2)}
\left(\frac{1}{4\pi \xi^2}\right)^{(3/2)}\times \nonumber \\ 
\int_0^\infty e^{-s}s^{(D-3)/2}e^{-|{\bf r}|^2/4s\xi^2}
\left(\frac{ds}{s}\right).
\label{3frac12}
\end{eqnarray}          

Let us consider the probability 
\begin{math} dP \end{math} that a Hydrogen atom nucleus  
is at a distance \begin{math} r\in dr  \end{math} away from 
an equilibrium position; 
i.e. \begin{math}dP(r;D)= 4\pi \rho_D(r)r^2dr \end{math}. 
The Debye-Waller (Gaussian) form factor corresponds to the formally 
infinite value of the dimensionality 
\begin{math} D_G=\infty  \end{math}. The fractal 
dimensionality associated with the secondary structure of 
\begin{math} \beta \end{math} sheets is 
\begin{math} D_\beta \approx 1.35  \end{math} while 
the dimensionality associated with the tertiary structure 
of global folding is 
\begin{math} D_{fold} \approx 1.68  \end{math}. Finally, 
the value of the fractal dimension (to be discussed below) 
of sets on which protons reside is 
\begin{math} D_p \approx 0.65  \end{math}. The plots 
in Fig.\ref{Fig2} of 
\begin{math}dP(r;D)/dr = 4\pi r^2\rho_D(r) \end{math} 
are shown for the above important values of 
\begin{math} D \end{math}. As the dimensionality 
\begin{math} D \end{math} is lowered (for {\em fixed} 
\begin{math} \left<|{\bf u}|^2\right> \end{math}), 
the distribution of particle positions are drawn inward    
toward the origin consistent with the definition of 
fractal dimensionality; i.e.  
\begin{math} P(r;D) \sim r^D \end{math} as 
\begin{math} r\to 0 \end{math}. The long ranged 
tail in \begin{math} dP(r;D)/dr \end{math} 
as \begin{math} r\to \infty  \end{math} is increased 
as \begin{math} D  \end{math} grows smaller.

\section{Experimental methods}

About  \begin{math} 600\ {\rm mg}\end{math} of Azurin powder was 
dehydrated under vacuum in a chamber in the presence of 
\begin{math} P_2O_5\end{math} for two days. It is 
known\cite{Gregory95} for globular proteins that 
the ratio \begin{math} h \end{math} of the weight 
of the water to the weight of the
protein in the sample in the range 
\begin{math} 0.35 \leq h \leq 0.40  \end{math}
corresponds to a one-shell hydration state.
A portion of Azurin powder was kept dry, at an hydration level estimated 
as \begin{math} h \leq 0,05 \{{\rm gm}(H_2O)/{\rm gm} (protein)\}\end{math}.
Another portion of powder was hydrated with heavy-water 
(\begin{math} D_2 O \end{math}).   
A level of 
\begin{math} h = 0.36 \{{\rm gm}(D_2O)/{\rm gm} (protein)\}\end{math} 
was prepared by controlled hydration
in a chamber under vacuum and in the presence of a saturated KCl 
heavy water solution. The water content was determined by measuring 
the increase in weight of the protein sample. In order to achieve an 
almost homogeneous hydration, the sample was arranged in the vacuum chamber 
to obtain the maximum exposure to the controlled environment. When the 
protein sample was exposed to the deuterium-hydrated environment, 
some of the protein
protons (hydrogen) were exchanged with environmental deuterons. It is 
known\cite{Settles96} that on a reasonable time-scale only a small and
definite proton-deuteron exchange takes place. To avoid an 
excess of proton-deuteron exchange, about one day was taken to reach 
the desired ratio $h$ starting from the dry powder. 
After each experimental run, both the 
\begin{math} D_2 O\end{math}--hydrated and the dry samples were  
weighed again in order to verify the stability of the hydration degree
throughout the experiment. In all the runs and for all the samples 
stability was at the level of   
\begin{math} (\Delta h/h)\approx 0.5 \% \end{math}.

Elastic neutron scattering scan were performed on the backscattering
spectrometer IN13  at the Institute Laue-Langevin. An energy 
resolution of \begin{math} \epsilon = 9 \mu{\rm eV}\end{math}, 
corresponding to an incident wavelength of 
\begin{math} 2.23 \end{math} \AA\ at a
backscattering angle of \begin{math} 3.3^o \end{math}, 
was achieved.  The rather high incident
energy of the thermal neutrons on IN13 allows an investigation of a wide 
range in \begin{math} Q^2 \end{math} in the interval 
( 0.08 \AA\begin{math}^{-2}<Q^2< 25.0 $\AA$^{-2} \end{math}). 

The measured quantity was the elastic part of the dynamic form 
factor\cite{Bee88},
\begin{equation} 
S({\bf Q},\omega )=\frac{1}{N}
\int_{-\infty}^\infty \left<\sum_{j=1}^N 
e^{-i{\bf Q\cdot r}_j(t)}e^{i{\bf Q\cdot r}_j(0)}\right>
e^{i\omega t}\left({dt\over 2\pi }\right),
\label{inelastic1}
\end{equation}
where \begin{math} {\bf r}_j(t) \end{math} is the position 
of the \begin{math} j^{th} \end{math} Hydrogen atom nucleus 
at time \begin{math} t \end{math}. In theory, the dynamic 
structure factor obeys the decomposition\cite{Bee88}
\begin{equation} 
S({\bf Q},\omega )=|F({\bf Q})|^2\delta (\omega )
+S^\prime ({\bf Q},\omega )
\label{inelastic2}
\end{equation}
into an elastic 
\begin{math} |F({\bf Q})|^2\delta (\omega ) \end{math}
and inelastic \begin{math} S^\prime ({\bf Q},\omega ) \end{math} 
part. In practise, one finds the elastic part by integrating the 
measured \begin{math} S({\bf Q},\omega ) \end{math} 
over a small but finite frequency bin 
\begin{math} \varpi \end{math}. 
For our case, the bin size was 
\begin{math} (\epsilon /e)=(\hbar \varpi /e)=9\ \mu{\rm Volts} \end{math}
so that  
\begin{equation}
\int_{|\omega | < \varpi } S({\bf Q},\omega ) d\omega \approx  
\left| \cal{F}({\bf Q})\right|^2 \ .
\end{equation}

The elastic scattering was measured for 3 hours in the temperature range 
\begin{math} 20\ {\rm K}<T< 300\ {\rm K} \end{math} 
(i) at 28 temperatures for 
\begin{math} D_2O \end{math}--hydrated and 
(ii) at 31 different temperatures for the dry powder.
In both cases, about 200 mg of sample was held in a
standard flat aluminum cell with internal spacing 
of .5 mm placed at an angle of 135$^o$ to the incident beam. 
The data were corrected to take into account the incident flux, 
cell scattering, self shielding and
the detector response which refers to the sample at the lowest temperature
(\begin{math}T_{min}=20\ {\rm K}\end{math}).  An average transmission 
probability of 0.95 was obtained. Neither multiple scattering nor 
multi-phonon corrections were applied. In both the hydrated and the 
dry cases, the scattering intensity was largely dominated
by the incoherent contribution of the protein hydrogen atoms (without 
proton-deuteron exchange in the hydrated sample). The coherent and other
incoherent contributions are estimated at less than the 5\% of the overall
scattering probability.

\section{Elastic neutron scattering data}

\begin{figure}[tp]
\scalebox {1.0}{\includegraphics[angle=270]{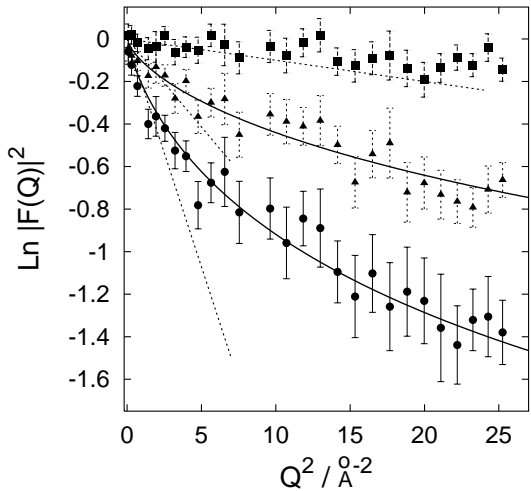}}
\caption{The experimental squared structure factor $|F(Q)|^2$ is shown 
for three temperatures ($50 K$, $215 K$ and $295 K$ from top to bottom) 
together with the theoretical form factor expressions (Eqs.(\ref{3frac5}) 
and (\ref{3frac5a}) for the fractal case, and Eq.(\ref{3frac0a}) for the 
Debye-Waller case). 
The solid lines are maximum likelihood fits to the 
fractal form factor. The dashed lines are fits to the
Debye-Waller form factor using experimental values of the mean square
displacement defined in Eq.(\ref{msd_definition}). Experimental data 
from dry Azurin powder (not shown) reveal a much
smaller departure from the Debye-Waller behavior.}
\label{Fig3}
\end{figure}

Shown in Fig.\ref{Fig3}  are measured values of $|F({\bf Q})|^2$ for the
hydrated Azurin powder obtained from neutron diffraction. 
The experimental results may be compared with (i) the standard Debye-Waller
model Eq.(\ref{3frac0a}) or (ii) the structure factor Eq.(\ref{3frac5}) 
of a set with fractal dimensionality \begin{math} D \end{math}. 
The dotted lines in Fig.\ref{Fig3} represent the
Debye-Waller form factor of Eq.(\ref{3frac0a}) in terms of the mean square
displacements. The mean square
displacement was obtained from the data in accordance with
Eq.(\ref{msd_definition}). In order to achieve an experimental definition
of the mathematical $Q \rightarrow 0$ limit we used a multiple fit
procedure. We chose the lowest 4, 5, 6 and 7 experimental points in $Q^2$.
For each temperature, the best fit was obtained as the reduced-$\chi^2$
weighted average of the four possibilities employing different numbers of
$Q^2$ values. The solid lines represent the maximum likelihood 
estimate of fractal dimension \begin{math} D \end{math} employed 
in Eq.(\ref{3frac5}). At very low temperatures, the description 
of the data in terms of the Debye-Waller form factor is reasonable. 
For dry Azurin (data not shown), only small deviations from the 
Debye-Waller form factor were observed. For hydrated Azurin, as the
temperature increases, very large deviations from the Debye-Waller appear
in the data. In regimes wherein these large deviations were observed, a
form factor of fractal dimension \begin{math} D \end{math} provided an 
accurate description of the experimental results.

In order to model the distances that the Hydrogen atom protons wander from
their mean equilibrium positions, we use a parametric Einstein model; 
i.e. the mean square displacement at temperature $T$ is described in 
terms of an effective temperature dependent Hooks force constant 
\begin{math} K(T) = m\Omega^2 (T)\end{math} where 
\begin{math} m \end{math} as the proton mass. 
\begin{equation}
\left<|{\bf u}|^2\right>_T = 
\left( \frac{3\hbar}{2m\Omega(T)} \right) 
\coth \left( \frac{\hbar\Omega(T)}{2k_B T} \right).
\label{Einst_model}
\end{equation}
If the Hydrogen atom proton were attached to an equilibrium position by 
a simple Hook's law spring \begin{math} K_0=m\omega_E^2 \end{math} to its 
equilibrium position, then a single Einstein frequency\cite{Lovesey84} 
\begin{math} \omega_E \end{math} would adequately fit
the experimental data for 
\begin{math} \left<|{\bf u}|^2\right>_T \end{math}. In reality, 
there is no single frequency \begin{math} \omega_E \end{math} 
for proton oscillations. Nevertheless, one may still define an effective
temperature dependent frequency \begin{math} \Omega(T) \end{math} to 
describe the mean strength of forces restoring the Hydrogen atom 
to its equilibrium position. In the hydrated Azurin protein samples, 
we find at the lowest temperatures
\begin{math} \Omega(T \rightarrow 0)=\omega_E \end{math} 
while at higher temperatures the restoring forces are softer and so 
the frequency is lower; i.e.   
\begin{math} \Omega(T > T_{dt})=\omega_S <\omega_E \end{math}. 
We have here introduced a dynamical transition temperature 
\begin{math} T_{dt}  \end{math} to characterize the softening of 
of the oscillation frequency spectrum as will now be discussed.

\begin{figure}[bp]
\scalebox {1.0}{\includegraphics[angle=270]{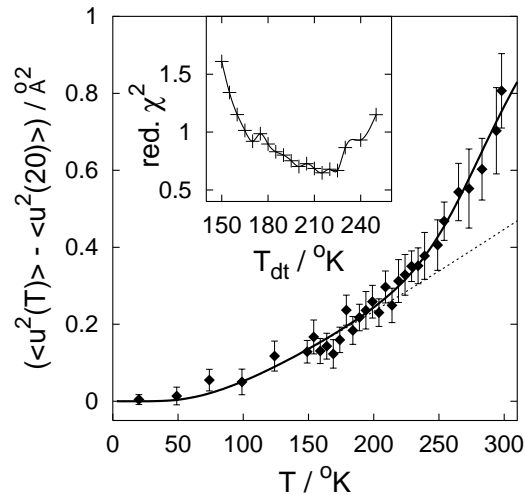}}
\caption{The mean square Hydrogen atom displacements of the hydrated
protein are obtained from a low-$Q^2$ analysis of $|F(Q)|^2$ as in
Eq.(\ref{msd_definition}).  The dashed line is the best least-square 
fit to the Einstein model Eq.(\ref{Einst_model}) with a frequency 
$\Omega(T)=\omega_E$ independent of temperature. 
The values of the reduced-$\chi^2$ associated with such fixed 
frequency fits are shown in the inset as 
a function of $T_{dt}$. The minimum reduced-$\chi^2$ is obtained
for $T_{dt} = 215\ K$ which corresponds to an Einstein harmonic frequency
$\omega_E/2\pi = 5.20 \pm 0.02\ THz$. In the case of the dry protein 
(data not shown), an Einstein model fit was possible for all temperatures 
with no dynamical transition. The solid curve is the best fit to the 
hydrated protein data with the effective frequency $\Omega(T)$ 
as a function of temperature.}
\label{Fig4}
\end{figure}

\begin{figure}[tp]
\scalebox {1.0}{\includegraphics{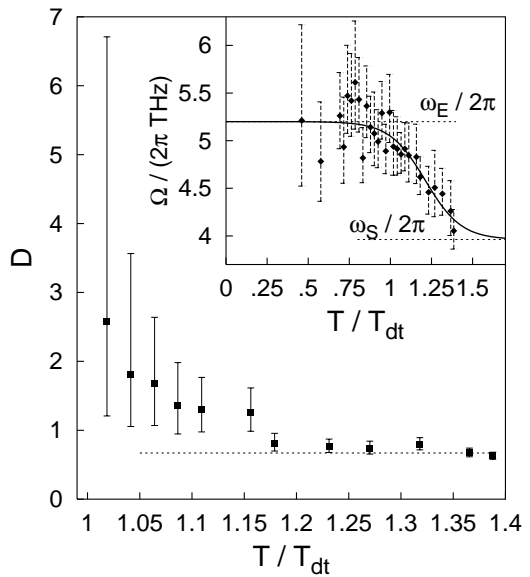}}
\caption{Shown is the fractal dimension $D$ obtained from the
maximum likelihood fit of hydrated protein experimental $|F(Q)|^2$
data to the fractal form factor of Eqs.(\ref{3frac5})
and (\ref{3frac5a}). The dashed
line corresponds to the asymptotic value of $D = 0.65 \pm 0.1$. The
dynamical transition temperature is $T_{dt} = 215 K$. For the dry
protein, likelihood convergence problems indicated that the 
fractal dimension $D$ was not a useful parameter for describing 
the data. In the insert, diamonds give the values of $\Omega(T)/2\pi$ 
in Eq.(\ref{Einst_model}) at each experimental temperature. The solid
line is the parameterized $\Omega(T)/2\pi$  obtained by fitting the
average mean-square displacements to Eq.(\ref{Einst_model}). The best
fit is obtained for $\omega_E/2\pi = (5.20 \pm 0.02) THz$ and
$\omega_S/2\pi = (3.96 \pm 0.1) THz$.}
\label{Fig5}
\end{figure}

The powder-averaged mean square displacement data from the hydrated sample
are shown in Fig.\ref{Fig4}. The data displays a marked temperature 
dependences not present in the dry Azurin sample. The marked 
temperature dependence 
has also been observed in many other types of hydrated protein samples 
employing many techniques
\cite{Parak71,Frauenfelder79,Smith91,Keller80,Parak81,Cohen81,Knapp82,
Rasmussen92} including neutron diffraction
\cite{Doster89,Smith91,Ferrand93,Lehnert93,Paciaroni00} in the temperature
range (\begin{math} 180\ {\rm K} <T_{dt}<220\ {\rm K} \end{math}). The 
behavior of the mean square displacements as 
a function of temperature is independent of secondary protein structure 
details, and has been characterized by a dynamical transition in
temperature from a harmonic to a non-harmonic regime. In order to
unambiguously determine a dynamical transition temperature 
\begin{math} T_{dt} \end{math}, we fit
a simple single frequency Einstein model to the experimental mean square
displacements for \begin{math} 0<T<T_{dt} \end{math} and then we 
varied the range values of \begin{math} T_{dt} \end{math}. 
The reduced-\begin{math} \chi(T_{dt})^2 \end{math} of these analyses 
was plotted as shown in the insert of Fig.\ref{Fig4}. The best 
reduced-\begin{math}\chi^2\end{math} agreement with
experimental data is obtained for 
\begin{math}T_{dt} = (215 \pm 5)\ {\rm K}\end{math}. The dotted
curve in the main of Fig.\ref{Fig4}  shows the relative theoretical fit to 
the data, corresponding to a frequency of 
\begin{math} (\omega_E / 2\pi) = 5.20 \pm 0.02\ {\rm THz} \end{math}
uniform in temperature. This curve reproduces well the experimental
behavior at low temperature taking quantum effects correctly into 
account. Above \begin{math} T_{dt} \end{math} such a simple Einstein 
model is clearly inadequate for the present hydrated sample. 
We fit the data with a smooth Fermi function merely to
extrapolate the temperature dependent 
\begin{math} \Omega(T) \end{math} from the low temperature
value of \begin{math}\Omega(T\rightarrow 0)=\omega_E \end{math} 
to the higher temperature value of 
\begin{math} \omega_S < \omega_E \end{math}. 
In the entire range of experimental temperatures, we have 
\begin{math} \omega_E \ge \Omega(T) \ge \omega_S\end{math}. 
The frequency \begin{math}\Omega(T) \end{math} is a smoothly 
decreasing function of temperature. The solid
curve in Fig.\ref{Fig4}  shows the parametric Einstein model fit to the 
experimental mean square displacement data as a function of 
temperature. The experimental softened frequency of the high 
temperature regime is given by
\begin{math} (\omega_S/2\pi)=3.96 \pm 0.1\ {\rm THz} \end{math}. 
In the inset of Fig.\ref{Fig5}, we show the
experimental points for the effective frequency $\Omega(T)$ as determined
by the experimental mean square displacements and Eq.(\ref{Einst_model}).
The solid curve represents the Fermi function fit to the experimental
$\Omega(T)$, the width of the transition being 
\begin{math} \Delta T = 25 \pm 1\ {\rm K}\end{math}. The temperature 
scale is in reduced units relative to the dynamical
transition temperature \begin{math} T_{dt} \end{math}.

Our central results concerning the fractal dimensionality of the hydrated
protein are shown in Fig.\ref{Fig5}. The maximum likelihood values of the fractal
dimension \begin{math} D \end{math} are plotted as function of the reduced 
temperature. Well above the dynamic transition temperature, where the 
effective restoring to equilibrium forces on the Hydrogen atom proton 
are softened, the fractal dimension of the set in which the proton resides 
is well defined and given by \begin{math}D=0.65 \pm 0.1 \end{math}. As 
the temperature is decreased, the dimension of the set grows until 
(for say \begin{math} D > 2 \end{math}) the likelihood fits are so 
broad as to become meaningless. In the dry case, there is no likelihood 
convergence in any temperature regime. Under hydrated conditions for  
\begin{math} T \geq 1.15\ T_{dt} \end{math}, the fractal 
dimension characterizes 
those protein samples whose parameters are neighboring the range 
allowing normal biological functions. 

\section{Conclusions}

Some remarks on our definition of fractal dimension are in order. For the
case under discussion, the fractal dimensionality $D$ is derived from
{\em incoherent} neutron scattering from single Hydrogen nuclei. For the
atomic length-scale of these experiments, $D$ is the fractal dimension of
the set on which a single proton resides. This fractal dimension $D$ is
defined independently of the positions of the other atoms. There has been
considerable previous and important work based on {\em coherent} scattering
experiments measuring fractal sets of dimension $d$ wherein $1.2 < d <
1.7$. Coherent scattering depends on amplitude superposition interference
from pairs of different atoms. The correlation dimension $d$, describes
sets that span a larger length scale and require the positions of atomic
pairs for their definition. These larger length scales are connected with
the pair-correlation properties of the $\alpha$-carbons of the polypeptide 
chain\cite{Wagner85,McCammon84} or with length-scales typical 
of the secondary and tertiary structures. The density fractal 
dimension \begin{math} D \end{math} of this work
describes the length scales of inter-atomic spacing while the correlation
fractal dimension $d$ describes the length scales of the full protein size.
(The length scale in diffraction experiments is $\sim Q^{-1}$.) To our
knowledge no relevant temperature dependence has been observed for the
correlation fractal dimension $d$. Only the interaction with environmental
conditions related to major changes (denaturation) in the global 
structure\cite{Chen86} had any effect on the values of 
\begin{math}d \end{math}. In the present case, the density fractal 
dimension \begin{math} D \end{math} is only well defined when 
the protein sample is at biologically relevant temperatures 
and in the biologically relevant hydrated condition. 
In regimes wherein the protein is in tact but not anywhere near the form 
required for normal biological life functioning, \begin{math} D \end{math} 
shows a strong divergence and becomes experimentally ill defined. 

A principle biological function of Azurin is electron transfer. It is
known that the electron-transfer efficiency of membrane protein
ensembles is strongly reduced by lowering the temperature below that of
the dynamical transition\cite{Parak80}. More recently, 
it has been shown \cite{Balabin00}
that protein atomic-level dynamics can amplify the electron transfer rates
via the quantum interference of amplitudes from alternative paths. The
regime of normal biological functions is also the regime for exhibiting
atomic length scale single particle sets of fractal dimension $D$. What
sense can be made of this observation?  

A large organization (protein) functions collectively. An individual
unit (atom) within the organization wanders to and fro. If the motions
of individual units involve a dimensionally large set (say two or three
dimensions), then the unit moves all over a neighborhood in very many
random directions and taking very many possible jumps of quite
indefinite lengths. If the unit moves in fewer directions and with a
smaller number of possible differing length scales, then the fractal
dimension \begin{math} D \end{math} of the set on which the unit lives 
is also smaller. This occurs above the dynamical transition temperature 
wherein the time scale (\begin{math} \Omega^{-1} \end{math})  
of the motion is increased. If each unit (atom) has 
fewer definite paths and fewer individual choices of
motion, then the resulting fractal dimensional residential set is
indicative of the healthy collective functioning of the larger
organization (protein).    

\section*{Acknowledgments}
DM would like to gratefully thank Alessandro Paciaroni for his relevant 
help in measurements and for useful discussions; Alessandro Desideri and 
Maria Elena Stroppolo are kindly acknowledged for having provided the 
samples; the ILL is acknowledged for experimental resources; this work 
was partially supported with an INFM grant.

\end{document}